\documentclass[conference]{IEEEtran}
\IEEEoverridecommandlockouts
\usepackage{cite}
\usepackage{amsmath,amssymb,amsfonts}
\usepackage{algorithmic}
\usepackage{algorithm}
\usepackage{graphicx}
\usepackage{textcomp}
\usepackage{xcolor}
\def\BibTeX{{\rm B\kern-.05em{\sc i\kern-.025em b}\kern-.08em
    T\kern-.1667em\lower.7ex\hbox{E}\kern-.125emX}}

\usepackage{makecell}
\usepackage{subfig}
\usepackage{bm}
\usepackage{amsthm}

\usepackage{tabularx}  
\usepackage{booktabs} 
    
\begin{document}

\title{Facet-offset Polytope Expansion for Dynamic Operating Envelopes of Aggregated Distribution Grid Resources
\thanks{The submitted manuscript has been created by UChicago Argonne, LLC, Operator of Argonne National Laboratory (“Argonne”). Argonne, a U.S. Department of Energy Office of Science laboratory, is operated under Contract No. DE-AC02-06CH11357. This work was supported in part by the U.S. Department of Energy, Office of Electricity. The U.S. Government retains for itself, and others acting on its behalf, a paid-up nonexclusive, irrevocable worldwide license in said article to reproduce, prepare derivative works, distribute copies to the public, and perform publicly and display publicly, by or on behalf of the Government.\\
($^*$Corresponding authors: Dr. Kumarsinh Jhala, kjhala@anl.gov, Feng Chen, feng.chen@colostate.edu).}
}

\author{\IEEEauthorblockN{Feng Chen$^{1,2,*}$, \textit{Student Member, IEEE}, Kumarsinh Jhala$^{1,*}$, \textit{Senior Member, IEEE}, \\Kiran Kumar Challa$^{1}$, \textit{Senior Member, IEEE}, Zongjie Wang$^{2}$, \textit{Senior Member, IEEE}}
$^{1}$Argonne National Laboratory, Lemont, IL, USA \hspace{2mm}
$^{2}$Colorado State University, Fort Collins, CO, USA
}
 
\maketitle

\begin{abstract}
Dynamic operating envelopes (DOEs) characterize the admissible power injection and withdrawal regions for distribution grid resources (DGRs) or aggregators while ensuring that distribution-system constraints remain satisfied. Convex relaxations and linearized power-flow models may fail to guarantee operational safety, while existing inner-approximation approaches are often limited by shape-specific conservativeness. This paper develops a facet-offset polytope expansion method (FOPE-sen) for constructing DOEs in three-phase unbalanced distribution systems. A two-step polytope expansion strategy is designed to enlarge the DOE volume and mitigate degeneracy caused by direction-order dependence. A sensitivity matrix is calculated to identify critical network constraints and guide facet-direction selection. Comprehensive numerical experiments demonstrate that the sensitivity-informed directions substantially improve feasible-region coverage over random directions. Furthermore, the proposed FOPE-sen method achieves greater coverage than a vertex-based expansion method while requiring fewer than 1/40 of its facets in the tested higher-dimensional cases.
\end{abstract}



\section{Introduction}

The increasing penetration of distribution grid resources (DGRs) is reshaping the operation of distribution systems. These resources can provide substantial flexibility to support grid operation and electricity markets. FERC Order No.~2222 encourages the participation of aggregated DGRs in wholesale electricity markets, creating new opportunities for DGRs to provide energy, capacity, and ancillary services. However, their widespread deployment also introduces new operational challenges \cite{JCA+26, JNP19}, particularly in three-phase unbalanced distribution networks, where voltage violations, line thermal overloads, and nonlinear AC power-flow effects can significantly limit the amount of DGR flexibility that can be safely accommodated \cite{xu2025robust, JPK+25}. Therefore, distribution system operators (DSOs) require computationally efficient and network-aware tools to quantify admissible DGR injection and withdrawal regions, ensure system reliability, better utilize existing network capacity, and delay costly grid upgrades.

Dynamic operating envelopes (DOEs) \cite{blackhall2020calculation} provide a promising mechanism for coordinating DGR operation and facilitating their secure participation in electricity markets. A DOE specifies the time-varying feasible range of DGR injections or withdrawals that can be accommodated by the network without violating operational constraints. By communicating these limits to DGRs or aggregators, DOEs can help DSOs maintain reliability while encouraging more market participation of DGRs. From an optimization perspective, computing DOEs can be interpreted as constructing an inner approximation of a nonconvex feasible region (FR). This raises two important questions. First, how can the feasibility of the solution be ensured for a nonconvex FR. Second, what shape should the inner approximation take.

Several methods have been developed to handle nonconvex FRs, including model approximation \cite{xu2025robust, liu2023robust}, Brouwer’s fixed-point theorem \cite{nguyen2018constructing}, and the Quotient Gradient System (QGS)-based approach \cite{zeng2025online}. Model approximation is the most widely adopted approach, in which the nonconvex FR is approximated using a convex relaxation or a linearized model. Then, the robust counterpart \cite{xu2025robust, liu2023robust} can be solved efficiently. Convex restriction \cite{nguyen2018constructing}, based on Brouwer’s fixed point theorem, uses bounding techniques to convexify the nonconvex constraints. This approach can obtain a guaranteed-feasible DOE for a nonconvex region, but it is often overly conservative. The QGS-based approach can efficiently find feasible solutions even from infeasible initial points, which are used for inner approximation expansion \cite{zeng2025online}. However, its applicability to three-phase unbalanced systems remains to be evaluated \cite{fang2026approximating}.


The shape design of DOEs is important for flexibility capture, communication efficiency, and scalability. In existing practice, DOEs are often represented by box-type limits \cite{liu2023robust}, where each DGR or aggregator receives independent upper and lower bounds. Box-type DOEs are suitable when there is no communication among DGRs. With aggregators, individual DGRs within a group may be managed more efficiently using other DOE shapes, such as zonotopes, hyperellipsoids \cite{liu2023robust}, and polytopes \cite{nguyen2018constructing, gong2024hosting}. Polytope-based envelopes offer the greatest flexibility because they can capture the coupling among DGR injections through a scalable set of linear inequalities while remaining compatible with downstream aggregator scheduling and electricity-market participation models. Reference \cite{gong2024hosting} expands a polytope-based envelope by iteratively adding new vertices, which changes both the number and the directions of the facets. However, the resulting redundant facets may limit the method’s scalability and increase the communication burden in high-dimensional problems.

\begin{figure}[t]
    \centering
    \includegraphics[width=0.99\linewidth]{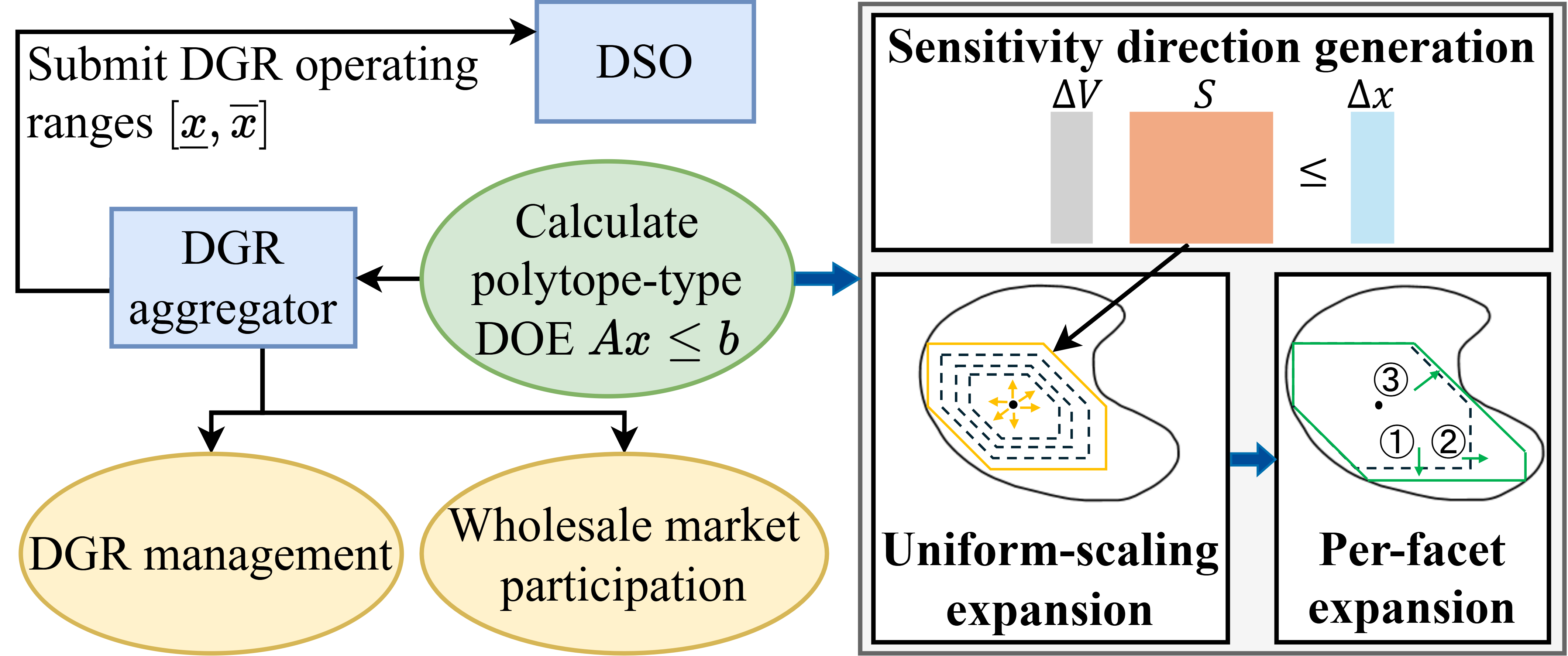}
    \caption{Facet-offset polytope expansion framework}
    \label{fig:framework}
\end{figure}

To avoid the approximation errors introduced by convex relaxations and linearized power-flow models while improving scalability in high-dimensional settings, this paper develops a facet-offset polytope expansion method for constructing network-aware DOEs in three-phase distribution systems, as illustrated in Fig.~\ref{fig:framework}. Rather than jointly optimizing the complete polytope representation \cite{gong2024hosting}, the proposed method first prespecifies the facet directions using DGR-based coordinate basis vectors (DGR directions) and sensitivity-informed network-critical directions. With these directions fixed, DOE construction is reduced to an offset-determination problem, namely, determining how far each facet can be moved outward from the expansion center along its prescribed direction. This problem is solved using a two-step facet-offset expansion procedure. The main contributions are summarized as follows:


\begin{itemize}
    \item A two-step facet-offset expansion approach (FOPE-sen) is developed to construct polytope-based DOEs with a controllable number of facets. Compared with the polyhedral vertex-based methods in \cite{gong2024hosting}, the controllable number of facets enables a more compact polytope representation, thereby facilitating downstream applications.
    
    \item A sensitivity-informed direction selection strategy is introduced to identify network-critical facet directions associated with voltage constraints. The facet location is determined by a two-step expansion strategy, which mitigates direction-order dependence while preserving the capability to explore the FR.
    
    \item Numerical experiments on the IEEE 123-bus test system demonstrate the effectiveness of the sensitivity-informed directions compared with random directions. Moreover, in tested higher-dimensional cases, the proposed method constructs larger DOEs than the vertex-based expansion method while requiring fewer than $1/40$ of its facets.
\end{itemize}






\section{Inner approximation of ACOPF feasible region}
A distribution system can be modeled as a network $\mathcal{G}(\mathcal{N}, \mathcal{E})$, where $\mathcal{N} = \{0, 1, \dots, |\mathcal{N}|\}$ denotes the set of buses and $\mathcal{E}$ denotes the set of lines. Define $\mathcal{N}^+ = \mathcal{N} \backslash \{0\}$, where the substation is indexed by 0. $i \sim j$ represents line $(i, j) \in \mathcal{E}$ or $(j, i) \in \mathcal{E}$. $\Phi_i$ and $\Phi_{ij}$ denote the phases of bus $i$ and line $i \sim j$, respectively. $V_i:=[V_i^{\phi}]_{\phi \in \Phi_i}, I_{ij}:=[I_{ij}^{\phi}]_{\phi \in \Phi_{ij}}, s_{i}:=[s_i^{\phi}]_{\phi \in \Phi_i}$ represent the complex voltage, complex line current, and complex injected power. $(\cdot)^H$ denotes the Hermitian transpose.

The bus injection model (BIM) \cite{gan2014convex} of power flows can be modeled as
\begin{equation} \label{for:power_flow}
    s_i = \sum_{j:i \sim j} \text{diag} \left[ V_i^{\Phi_{ij}}(V_i^{\Phi_{ij}} - V_j^{\Phi_{ij}})^H y_{ij}^H \right]^{\Phi_i}, ~ i \in \mathcal{N}.
\end{equation}

There are some physical and operating constraints:
\begin{subequations} \label{for:OPF_oper_consts}
    \begin{align} 
         & s_i \in \mathcal{S}_i, ~ i\in \mathcal{N}^+, \label{for:s_limits} \\
         & V_0 = V_0^{\text{ref}}, \label{for:V_ref} \\
         & \underline{V}_i^{\phi} \leq |V_i^{\phi}| \leq \overline{V}_i^{\phi}, ~ i \in \mathcal{N}^+, \phi \in \Phi_i, \label{for:V_limits} \\
         & |I_{ij}^{\phi}| \leq \overline{I}_{ij}^{\phi}, ~ i \sim j, \phi \in \Phi_{ij}, \label{for:I_limits}
     \end{align}
\end{subequations}
where the substation voltage is fixed at magnitude $|V_0^{ref}|=1$ p.u. with balanced phase angles in \eqref{for:V_ref}. The definition of $\mathcal{S}_i$ depends on the type of unit connected to bus $i$. In this paper, we assume
$\mathcal{S}_i = \{s_i | \underline{P}_i \leq \text{Re}(s_i) \leq \overline{P}_i, \underline{Q}_i \leq \text{Im}(s_i) \leq \overline{Q}_i\}$,
where $s_i^{\phi} = P_i^{\phi} + j Q_i^{\phi}$. If no unit is connected to bus $i$, then $S_i=\{0\}$. For a fixed load, $\mathcal{S}_i$ contains only its fixed complex-power vector. Let $x$ denote the DGR power injections as defined in \eqref{def:x}. The DOE is then the projection of the ACOPF FR onto the DGR injection space, as defined in \eqref{def:DOE}.
\begin{equation} \label{def:x}
    x = [P_i^{\phi}, Q_i^{\phi}, \forall i \in \mathcal{N}^{DGR}, \forall \phi \in \Phi_i^{DGR}]^\top
\end{equation}

\begin{equation} \label{def:DOE}
    \mathcal{F}: = \{x \mid \exists V_i, I_{ij}, ~\text{s.t.} ~ \eqref{for:power_flow}, \eqref{for:OPF_oper_consts} \}.
\end{equation}

For a three-phase unbalanced system, $\mathcal{F}$ is usually a nonconvex set. In traditional practice, DOE is modeled as a box $\mathcal{B}:=\{x \mid  \underline{x} \leq x \leq \overline{x} \}$ to remove the dependence between different DGRs. When DGRs are collectively managed by an aggregator, a polytope-based DOE can enable more efficient coordination among them. This paper aims to construct a maximum-volume polytope-based DOE $\mathcal{P}$ that preserves the security of the distribution system, as shown in \eqref{pro:max_poly_vol}.
\begin{subequations} \label{pro:max_poly_vol}
    \begin{align}
        \max_{A,b} \quad & \text{Vol}(\mathcal{P}) \label{pro:max_poly_vol_obj} \\
        \text{s.t.} \quad & \mathcal{P}: = \{ x\in \mathbb{R}^{m} \mid Ax \leq b \} \subseteq \mathcal{F}, \label{pro:max_poly_consts}
    \end{align}
\end{subequations}
where $A \in \mathbb{R}^{n\times m}$ and $b \in \mathbb{R}^n$. In \eqref{pro:max_poly_consts}, every row $a_r^{\top} x \leq b_r$ denotes a hyperplane, where $a_r$ represents the hyperplane's direction. Determining optimal $A$ and $b$ simultaneously is difficult, since there are many bilinear terms in $Ax$ \cite{castro2015tightening}. Instead of directly solving \eqref{pro:max_poly_vol}, iterative methods are applied to determine $A$ and $b$ automatically, but the efficiency will be a bottleneck in high-dimensional problems if the number of hyperplanes is not managed suitably \cite{gong2024hosting, deits2015computing}. In our design, we first fix $A$ according to DGR coordinate directions and the voltage sensitivity matrix, then the problem is reduced to finding an optimal $b$ for a predetermined $A$.

\section{Facet-offset polytope expansion} \label{sec:poly_expand}
\subsection{Two-step facet-offset expansion based on bisection methods}


Assuming there is an expansion center $x_0 \in \mathcal{F}$ and several predetermined facet directions $a_r, r = 1,2,\dots,n$. In practice, the center can be selected as a representative operating point based on state estimation or historical data. In this paper, a set of operating points within $\mathcal{B}$ is randomly generated, from which the feasible point with the largest total distance to the FR boundary is selected as the expansion center.The procedure to determine the facet directions is introduced in Sec.~\ref{sec:sens}.

We employ a one-dimensional bisection search to determine the offset of each facet, thereby avoiding the excessive computational burden and potential numerical instability associated with directly solving the ACOPF problem. The oracle used to judge the feasibility of the PF problem is defined as follows: 
\begin{equation} \label{eq:oracle}
    \mathcal{O}(x) =
    \begin{cases}
        1, & \text{if } x \in \mathcal{F},\\
        0, & \text{otherwise}.
    \end{cases}
\end{equation}

In the one-dimensional search along direction $a_r$,
\begin{equation} \label{for:one_dim_search}
    x = x_0 + d_r \cdot a_r,
\end{equation}
where $d_r \geq 0$ is the scalar offset. After several bisection iterations, the maximum admissible offset $L_r$ defined in \eqref{pro:L_r} can be obtained within a prescribed numerical tolerance.
\begin{subequations} \label{pro:L_r}
    \begin{align}
        L_r: = \max \quad & d_r \\
        \text{s.t.} \quad & \eqref{for:one_dim_search}, x \in \mathcal{F}.
    \end{align}
\end{subequations}
After we obtain $L_r$, the maximum facet offset can be determined by a two-step method: uniform-scaling expansion \eqref{pro:plane_loc_phase1} and per-facet expansion \eqref{pro:plane_loc_phase2}.
\begin{subequations} \label{pro:plane_loc_phase1}
    \begin{align}
        s^* = \max \quad & s \\
        \text{s.t.} \quad & s \in [0, 1], \\
        & v \in \mathcal{F}, ~\forall v \in \text{H2V} (Ax - Ax_0 \leq sL),
    \end{align}
\end{subequations}
where $\operatorname{H2V}(\cdot)$ returns the vertex set of the polytope defined by its hyperplane-based H-representation. During the uniform-scaling step, the polytope preserves its shape, while its size is controlled by a scalar $s$.
\begin{subequations} \label{pro:plane_loc_phase2}
    \begin{align}
        d_r^* = \max \quad & d_r \\
        \text{s.t.} \quad &  d_r \in [s^* L_r, L_r], \\
        & v \in \mathcal{F}, ~\forall v \in \text{H2V} (Ax - Ax_0 \leq d). \label{pro:plane_loc_phase2_3}
    \end{align}
\end{subequations}

In \eqref{pro:plane_loc_phase2_3}, note that all the elements in $d$ except $d_r$ are fixed at their initial values or the values in the last iteration with the bisection search. The solution $d_r^*$ in \eqref{pro:plane_loc_phase2} leads to a hyperplane $a_r^\top x \leq b_r^*$ in the final polytope, where $b_r^*=a_r^{\top} x_0 + d_r^*$. The bisection method is applied to obtain the solutions of models \eqref{pro:L_r}--\eqref{pro:plane_loc_phase2}, which is summarized in Algorithm 1.

For nonconvex AC feasible regions, vertex checks and zero observed sampling error provide empirical validation, not a formal safety guarantee for the entire envelope. Smaller expansion step size and denser sampling can improve the treatment of nonconvexity, but significantly increase the computational burden, particularly in high-dimensional settings. Stronger certification remains future work.


\renewcommand{\algorithmicrequire}{\textbf{Input:}}
\renewcommand{\algorithmicensure}{\textbf{Output:}}







\begin{algorithm}
\caption{Two-step facet-offset polytope expansion}
\label{alg:twophase}
\begin{algorithmic}[1]
\REQUIRE Search directions $A = [a_1,\ldots,a_n]^\top$, PF feasibility oracle $\mathcal{O}(\cdot)$, expansion center $x_0$,
$\{L_r\}_{r=1}^n$, tolerance $\epsilon$, DGR power range $[\underline{x},\overline{x}]$
\ENSURE offsets $\{d_r^*\}_{r=1}^n$ with a vertex-checked polytope $\mathcal{P}(d^*)=\{ x:A x\le A x_0+d^*\}\cap[\underline{x},\overline{x}]$
\vspace{2pt}
\STATE \textbf{Phase 1: uniform-scaling expansion by solving \eqref{pro:plane_loc_phase1}}
\STATE $L_{\max}\leftarrow \max_r L_r$;\quad $s_{lo}\leftarrow 0$;\quad $s_{hi}\leftarrow 1$
\WHILE{$s_{hi}-s_{lo} > \epsilon/L_{\max}$}
    \STATE $s\leftarrow \tfrac12(s_{lo}+s_{hi})$
    \IF{ $v \in \mathcal{F}, \forall v \in \text{H2V}(Ax - Ax_0 \leq sL)$ }
        \STATE $s_{lo}\leftarrow s$
    \ELSE
        \STATE $s_{hi}\leftarrow s$
    \ENDIF
\ENDWHILE
\STATE $d\leftarrow s_{lo} L$
\STATE \textbf{Phase 2: per-facet expansion by solving \eqref{pro:plane_loc_phase2}}
\REPEAT
    \STATE $\Delta_{\max}\leftarrow 0$
    \FOR{$r=1$ \TO $n$}
        \STATE $p_{lo}\leftarrow d_r$;\quad $p_{hi}\leftarrow L_r$;\quad $d_r^{0}\leftarrow d_r$
        \WHILE{$p_{hi}-p_{lo}>\epsilon$}
            \STATE $d_r \leftarrow \frac{1}{2}(p_{lo}+p_{hi})$
            \IF{ $v \in \mathcal{F}, \forall v \in \text{H2V}(Ax - Ax_0 \leq d)$ }
                \STATE $p_{lo}\leftarrow d_r$
            \ELSE
                \STATE $p_{hi}\leftarrow d_r$
            \ENDIF
        \ENDWHILE
        \STATE $d_r\leftarrow p_{lo}$;\quad $\Delta_{\max}\leftarrow\max(\Delta_{\max},\,d_r-d_r^{0})$
    \ENDFOR
\UNTIL{$\Delta_{\max}<\epsilon$}
\RETURN $\{d_r\}_{r=1}^n$
\end{algorithmic}
\end{algorithm}

\subsection{Sensitivity-based direction selection} \label{sec:sens}
Choosing reasonable facet directions under a facet number limit is important for algorithm efficiency and approximation accuracy. An intuitive type of direction is DGR coordinate directions $\{\pm e_i\}_{i=1}^{m}$, where $e_i \in \mathbb{R}^{m}$ denotes the $i$-th standard basis vector. Polytope expansion under this type of direction aims to maximize DGR power injection ranges.


In addition to the DGR coordinate directions, we use sensitivity-based directions for each monitored quantity $g_j(\cdot)$, such as voltage, current, or line power, as defined in \eqref{eq:sens}.
\begin{equation}
  \label{eq:sens}
  \Delta g \approx S \Delta x,\qquad S_{jk} = \left. \frac{\partial g_j}{\partial x_k} \right|_{x_0},
\end{equation}
where the sensitivity matrix $S$ is evaluated at the selected center $x_0$. It can be numerically calculated by $S_{jk} = \left( g_j(x_0 + \Delta \cdot e_k) - g_j(x_0 - \Delta \cdot e_k) \right) / 2\Delta$, where $\Delta$ is a small scalar and $g_j(\cdot)$ is evaluated from the power-flow solution obtained during the oracle call $\mathcal{O}$ in \eqref{eq:oracle}. The gradient $s_j = S_{j, :}^\top$ represents the direction of the fastest increase of the scalar function $g_j(\cdot)$. So $\pm s_j$ are two directions that can move $g_j(x_0)$ to its bounds quickly under the static assumption of $S$, where the shortest distance from the center to the boundary (headroom) can be calculated by
\begin{equation}
    \rho_j^+ = \frac{\overline{g}_j - g_j(x_0)}{\| s_j \|_2}, \quad \rho_j^- = \frac{g_j(x_0) - \underline{g}_j}{\| s_j \|_2}.
\end{equation}

Note that the sensitivity matrix $S$ may change nonlinearly under different operating points \cite{JNP18}. However, headroom $\rho^+$ and $\rho^-$ give us important guidance to determine the critical facet directions. The smaller the headroom, the more likely the constraint $ g_j (\cdot) $ is to reach its limit. After ranking all the values of headroom, we can select the normalized directions $\pm u_j = \pm s_j/\| s_j \|$ corresponding to the $N_s$ smallest headroom values, which is denoted by $S_{\text{top}(N_s)}$. To avoid near-parallel directions, a candidate $u_k$ is discarded if any selected $u_j$ satisfy $|u_j^\top u_k| > \tau$, where $\tau$ is a cosine threshold. The voltage sensitivity matrix is used to determine the sensitivity-informed direction.

Finally, the resulting $A$ in \eqref{eq:dir_select} is determined by DGR coordinate directions and sensitivity directions, which is held fixed throughout both expansion stages.
\begin{equation} \label{eq:dir_select}
    A = \begin{bmatrix}
        A_{DGR}\\ A_{sen}
    \end{bmatrix} = \begin{bmatrix}
        [e_1, -e_1, \dots, e_m, -e_m]^\top \\
        S_{\text{top}(N_s)},
    \end{bmatrix}_{(2m+N_s) \times m}
\end{equation}

\subsection{Performance evaluation} \label{sec:metric}
Since the AC power flow (ACPF) FR is nonconvex, the convex polytope may include some infeasible solutions. It's important to evaluate the approximation rate and safety of the final polytope. Therefore, two metrics are designed as 
\begin{subequations} \label{for:metrics}
    \begin{align}
        \text{Coverage Rate (CR)} &:=
        \frac{N_{\text{poly,fea}}}{N_{\text{AC,fea}}}
        \label{for:metrics_MR}, \\
        \text{Error Rate (ER)} &:=
        \frac{N_{\text{poly,in}}}
        {N_{\text{poly,fea}} + N_{\text{poly,in}}}
        \label{for:metrics_ER}.
    \end{align}
\end{subequations}
where $N_{\text{AC,fea}}$ counts the feasible samples.
$N_{\text{poly,fea}}$ and $N_{\text{poly,in}}$ count the feasible and infeasible samples that lie within the
calculated polytope, respectively.

Since the sampling-based evaluation above does not scale well to high-dimensional problems, the volume of the maximum-volume ellipsoid inscribed in the constructed polytope is used as a scalable proxy for the envelope size.

\section{Numerical experiments} \label{sec:case}
The IEEE 123-bus distribution system is used for the experiments. Three DGRs, denoted by G1, G2, and G3, are connected to nodes 300a, 114a, and 48a, respectively. Each DGR has an active-power range of $[0,200]$ kW and a reactive power range of $[-100, 100]$ kVAr. The power injections of these DGRs are denoted by P$i$ and Q$i, i=1,2,3$. The feasibility oracle $\mathcal{O}$ is implemented using PowerModelsDistribution.jl \cite{fobes2020powermodelsdistribution}, with feasibility assessed under voltage limits of $[0.95, 1.05]$ p.u. The expansion center is selected from 500 random samples. Parameter $\epsilon$ in Alg.~\ref{alg:twophase} is set to 1 kW/kVAr.




Three approaches are compared to evaluate our method:


\begin{itemize}
    \item Facet-Offset Polytope Expansion with sensitivity-informed facet directions (FOPE-sen): the proposed method in Sec.~\ref{sec:poly_expand} with facet directions in \eqref{eq:dir_select}.
    
    \item FOPE-rnd: this variant is identical to FOPE-sen, except that the sensitivity-informed facet directions are replaced with an equal number of randomly generated directions.
    
    \item FOPE-box: this variant uses only DGR coordinate directions in \eqref{eq:dir_select}, resulting in a hyperrectangle.
\end{itemize}

\subsection{Expansion process}
\begin{figure}[htbp]
\vspace{-10pt}
\centering
\subfloat[Uniform-scaling expansion]{\includegraphics[width=0.42\linewidth]{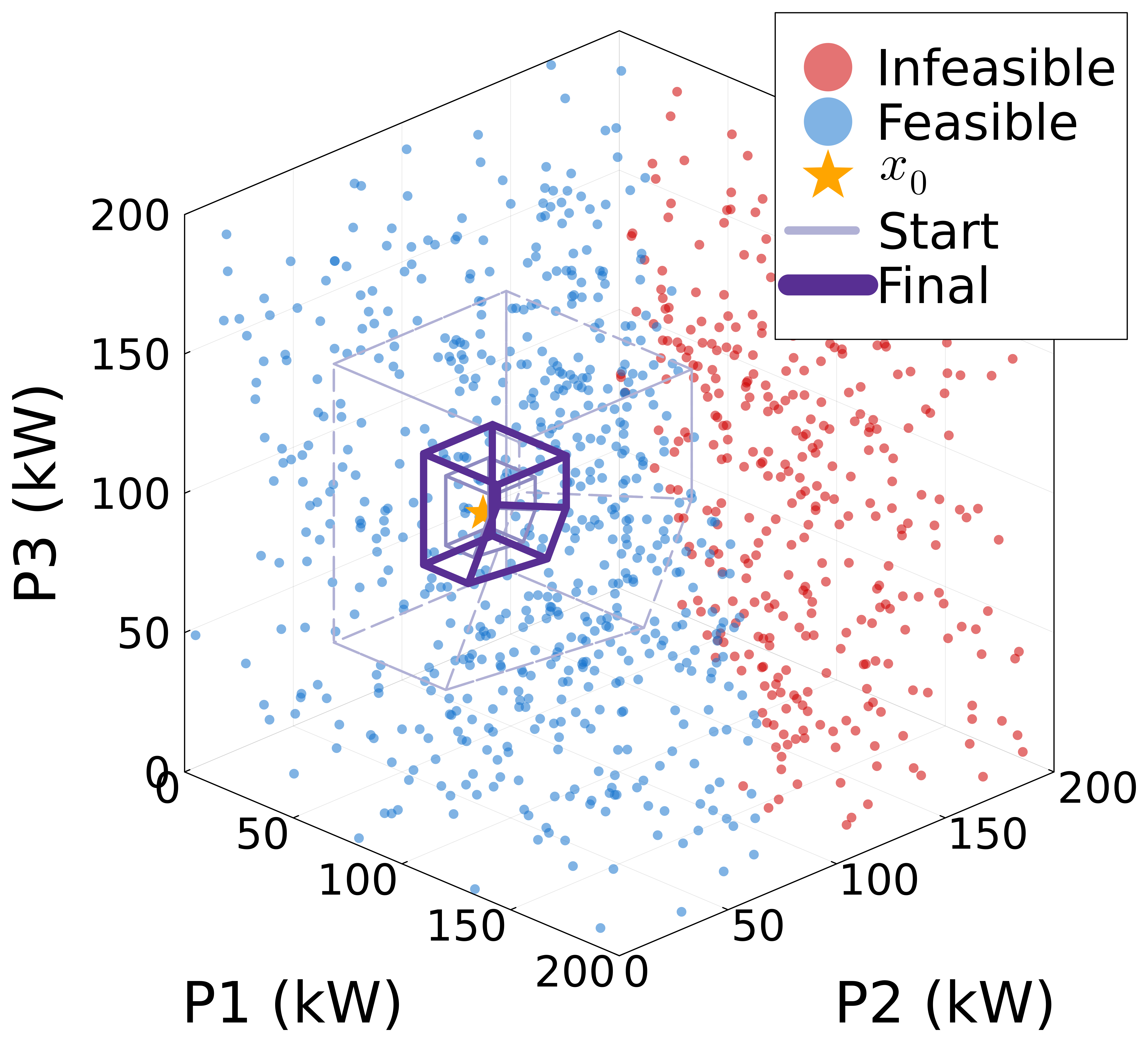}
\label{fig:3D_proj_FOPE-rdn_phase1}}
\hfil
\subfloat[Per-facet expansion]{\includegraphics[width=0.42\linewidth]{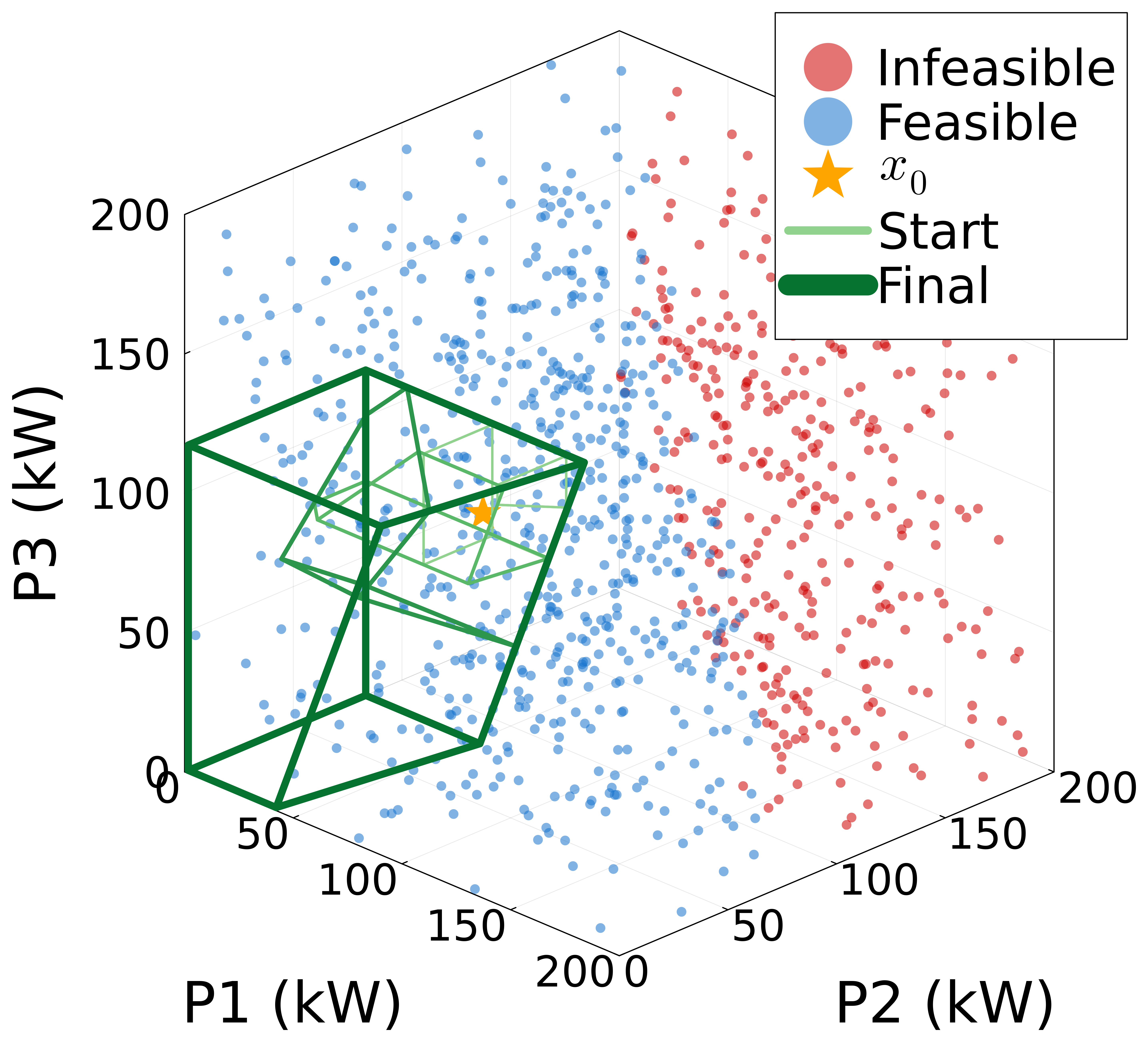}
\label{fig:3D_proj_FOPE-rdn_phase2}}
\caption{Slice of the DOE constructed using FOPE-rnd in the P1-P2-P3 space with fixed Q.}
\label{fig:3D_proj_FOPE-rdn}
\end{figure}

The polytope expansion process of FOPE-rnd is illustrated in Fig.~\ref{fig:3D_proj_FOPE-rdn}. For visualization, the reactive powers are fixed at the expansion center. In the uniform-scaling step, the polytope expands while preserving its shape within the prescribed power-limit box. In the per-facet expansion step, each facet is sequentially shifted outward along its associated direction whenever the resulting polytope remains feasible. This second step recovers additional flexibility that cannot be captured by uniform scaling alone due to the fixed polytope shape.

\begin{figure}[htbp]
\vspace{-10pt}
\centering
\subfloat[Uniform-scaling expansion]{\includegraphics[width=0.42\linewidth]{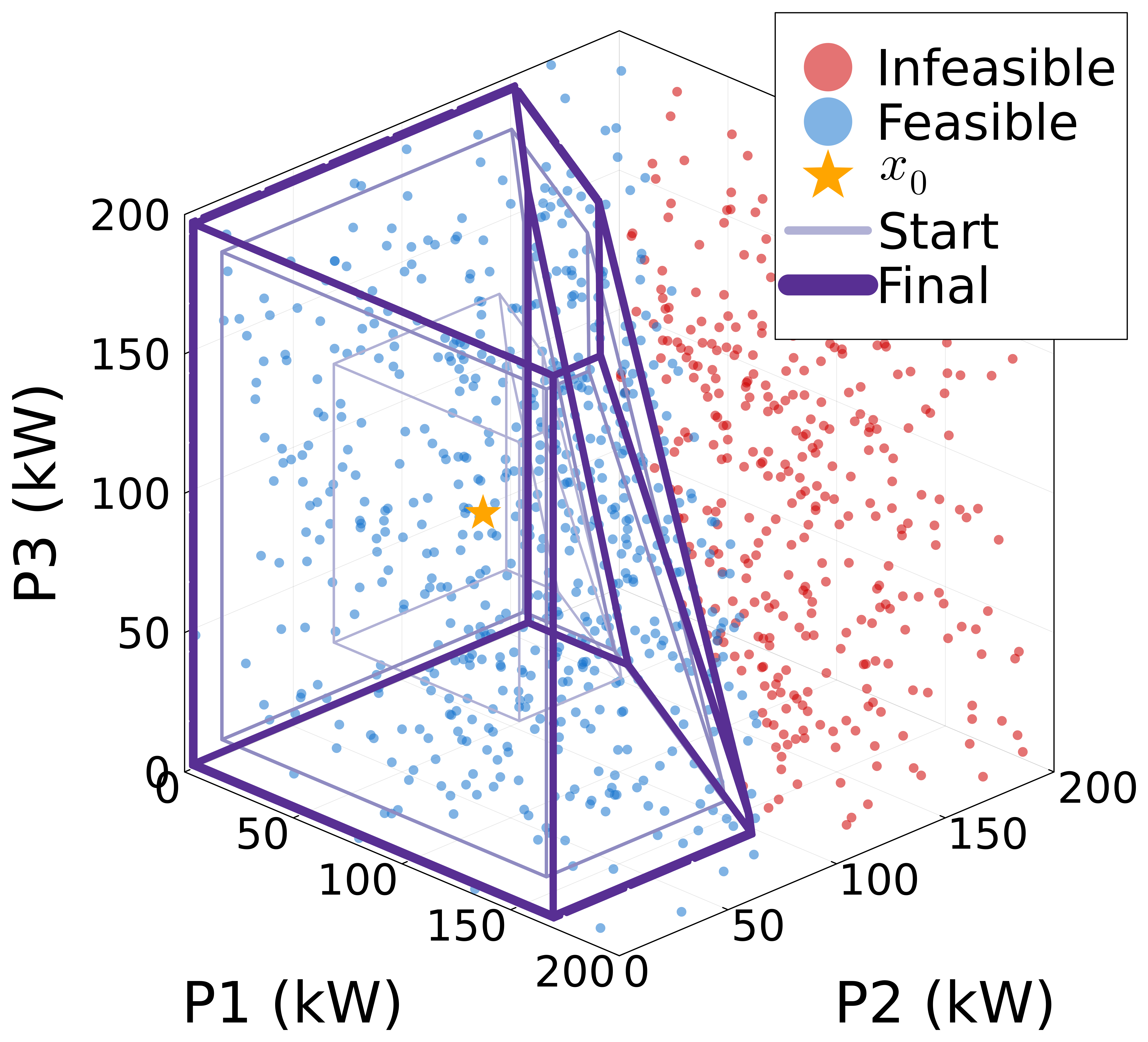}
\label{fig:3D_proj_FOPE-sen_phase1}}
\hfil
\subfloat[Per-facet expansion]{\includegraphics[width=0.42\linewidth]{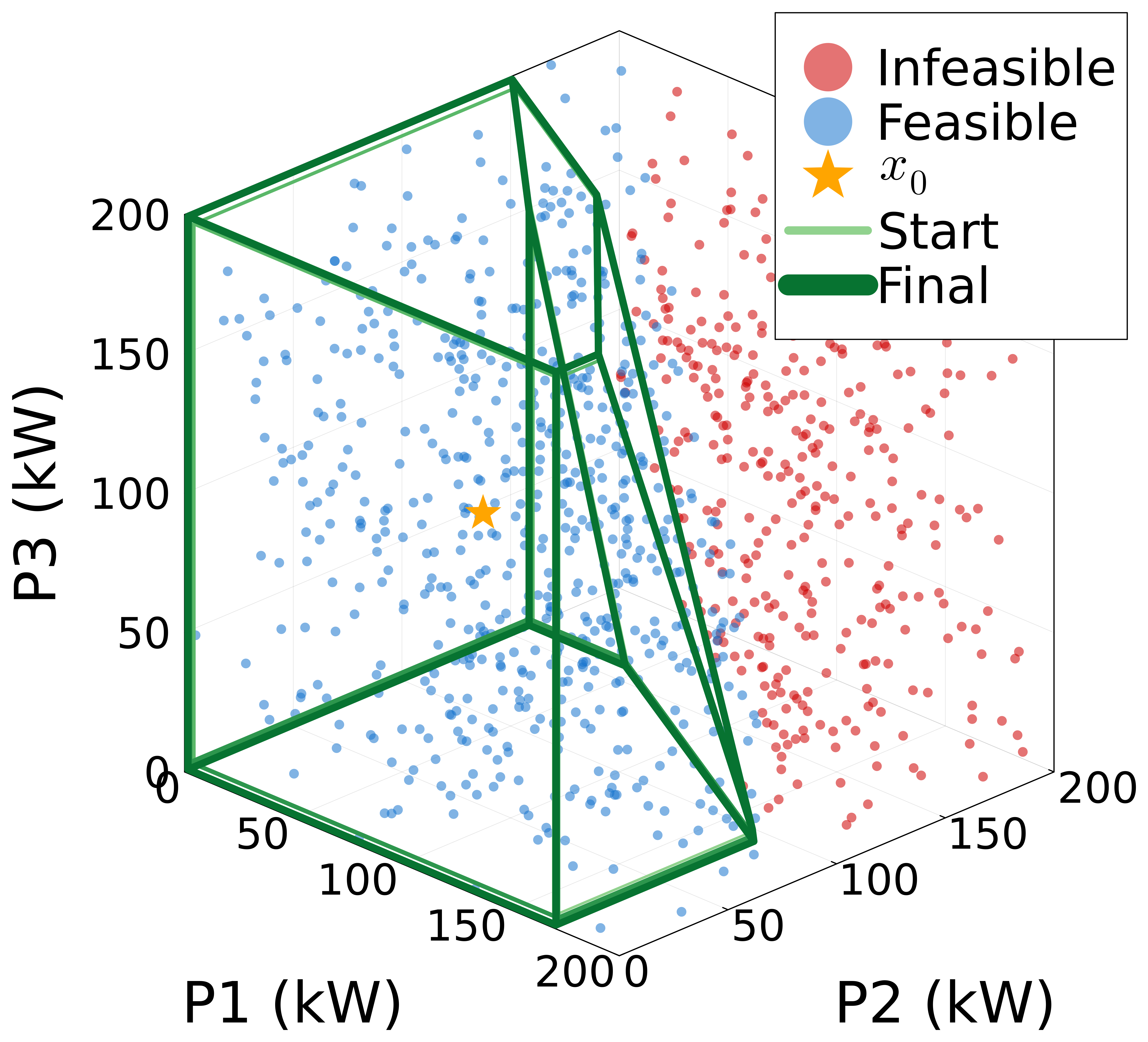}
\label{fig:3D_proj_FOPE-sen_phase2}}
\caption{Slice of the DOE constructed using FOPE-sen in the P1-P2-P3 space with fixed Q.}
\label{fig:3D_proj_FOPE-sen}
\end{figure}

The polytope expansion process of FOPE-sen is similar to that of FOPE-rnd, as illustrated in Fig.~\ref{fig:3D_proj_FOPE-sen}. However, only limited facet movement—from the light-green polytope to the dark-green polytope—is observed during the per-facet expansion in Fig.~\ref{fig:3D_proj_FOPE-sen_phase2}. This indicates that the FR representable by the prescribed facet directions is largely captured during the uniform-scaling step.


\subsection{Performance metric comparison across FOPE variants}
Performance metrics are summarized in Table~\ref{tab:method_comparison}, where Dia. denotes the diameter of the maximum-volume inscribed ellipsoid of the constructed polytope, while Range/dim. denotes the $m$-th root of the ellipsoid volume, where $m$ is the problem dimension. CR and ER are estimated from 10,000 samples drawn from $\mathcal{B}$. Relatively low coverage in each dimension leads to small CR values for FOPE-box and FOPE-rnd in high-dimensional spaces, e.g., $(40.8/200)^6 \approx 7.2 \times 10^{-5}$. For FOPE-sen, the preparation, uniform expansion, and per-facet expansion stages took 95.4, 58.0, and 84.8 s, respectively.

Two key observations can be drawn. First, the choice of facet directions has a significant impact on FR coverage, as indicated by the relatively large standard deviations observed for FOPE-rnd. Second, the polytope constructed by FOPE-sen achieves a substantially larger volume than those obtained by FOPE-box and FOPE-rnd, demonstrating the effectiveness of the sensitivity-informed direction enhancement. Note that only three sensitivity-based facets are retained in this case, as near-parallel directions are discarded with a threshold $\tau = 0.98$.

\begin{table}[t]
\centering
\caption{Comparisons for different expansion methods.}
\label{tab:method_comparison}
\setlength{\tabcolsep}{3.2pt}
\begin{tabular}{lcccccc}
\hline
Method & \makecell[c]{Facet \\ num.} & \makecell[c]{Range/dim. \\ (kW/kVAr)} & \makecell[c]{Dia. range \\(kW/kVAr)} & \makecell[c]{Running \\ time (s)} & \makecell[c]{ACPF \\ calls}& \makecell[c]{CR \\ (\%)} \\
\hline
FOPE-box & 12 & 40.8 & [22.9, 115.5] & 133.7 & 2132 & 0.32 \\
\makecell{FOPE-\\ rnd$^*$} & 15 & \makecell[c]{$37.2$ \\ $\pm 7.5$} & \makecell[c]{$[28.1 \pm 9.3,$ \\ $ 108.3 \pm 19.4]$} & \makecell[c]{$248.0$ \\ $\pm 41.6$} & \makecell[c]{$3155$ \\ $\pm 220$} & \makecell[c]{$0.34$ \\ $\pm 0.29$} \\

FOPE-sen & 15 & 92.2 & [70.3, 200.6] & 238.2 & 3511 & 43.75 \\
\hline
\end{tabular}
\begin{minipage}{0.98\columnwidth}
\footnotesize
\raggedright
Note: $^*$ indicates statistical results are from five independent trials. ER is zero for all cases.
\end{minipage}
\end{table}

\subsection{Robustness and scalability}
To evaluate the robustness and scalability of the proposed FOPE-sen method, we compare it with the Vertex-based Polytope Expansion (VBPE) method \cite{gong2024hosting}. At each iteration, the convex hull must be recomputed after adding a new vertex, which limits the scalability of VBPE in high-dimensional spaces. Accordingly, we terminate VBPE when either $\binom{\text{v}}{m} \leq 3\times 10^7$
\cite{mcmullen1970maximum} or the computation time exceeds one hour, where v denotes the number of vertices. The early-stage assessment proposed in \cite{gong2024hosting} is not adopted, as it may lead to rapid growth in the number of vertices in high-dimensional settings.

\begin{figure}[htbp]
\vspace{-10pt}
\centering
\subfloat[VBPE (v=52, m=6)]{\includegraphics[width=0.45\linewidth]{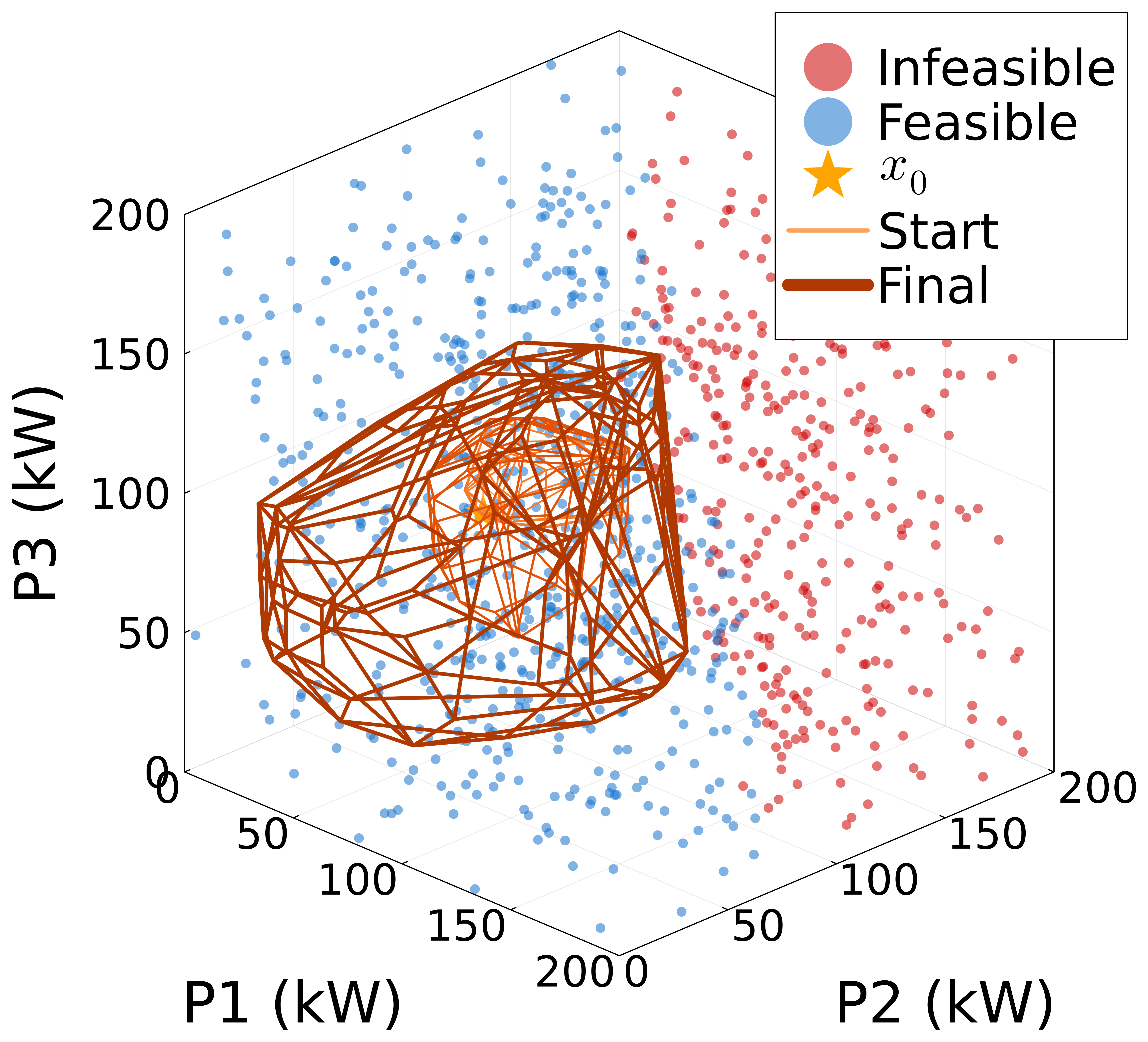}
\label{fig:VBPE}}
\hfil
\subfloat[VBPE vs. FOPE-sen]{\includegraphics[width=0.52\linewidth]{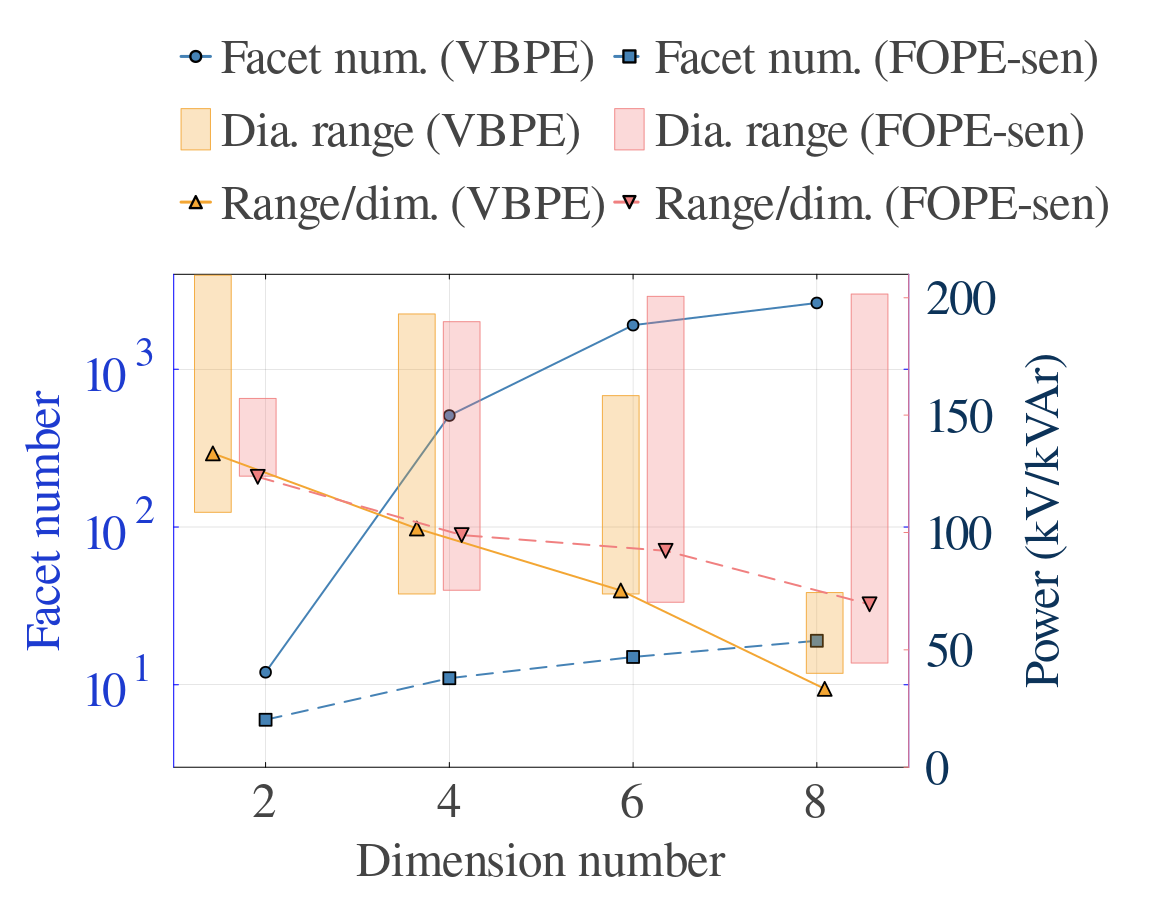} \label{fig:comp}}
\caption{DOE construction using VBPE and simulation results with different dimensions.}
\label{fig:VBPE&comp}
\end{figure}

In Fig.~\ref{fig:VBPE}, the DOE constructed by VBPE contains more facets but has a smaller volume than that obtained by FOPE-sen in Fig.~\ref{fig:3D_proj_FOPE-sen_phase2}. The performance of VBPE and FOPE-sen across different dimensions is compared in Fig.~\ref{fig:comp}. VBPE can provide better FR coverage with $m=2$ because it continuously increases the number of vertices and facets to approximate the original FR. However, when $m \geq 4$, FOPE-sen captures a comparable or even larger portion of the FR using fewer than $1/40$ of the facets required by VBPE. This result demonstrates the robustness and scalability of the proposed method in tested higher-dimensional spaces. The underperformance of VBPE under limited computational resources is mainly attributed to the sparsity and low information efficiency of the generated vertices, which leads to a bad-shaped polytope.




\section{Conclusions}
This paper presented a facet-offset polytope expansion framework with sensitivity-informed directions (FOPE-sen) for constructing DOEs in three-phase unbalanced distribution systems. The facet directions are selected from DGR coordinate directions and sensitivity-informed network-critical directions, while the facet locations are determined through a two-step bisection procedure guided by a power-flow feasibility oracle. Numerical results demonstrate that FOPE-sen captures at least 106\% more flexibility per dimension than box-shaped and random-direction expansion methods. Moreover, FOPE-sen constructs DOEs comparable to or larger than those of VBPE while using fewer than 1/40 as many facets in dimensions greater than four. Future work will focus on stronger safety verification for nonconvex AC feasible regions, scalability in high-dimensional spaces, and extensions to multi-aggregator allocation and market-participation applications.



\bibliographystyle{IEEEtran}
\bibliography{bibtex/bib/myRef}{}

\end{document}